\documentclass{article}

\usepackage[utf8]{inputenc}

\usepackage{authblk}

\usepackage[a4paper, total={6in, 10in}]{geometry}

\usepackage{graphicx}

\usepackage{amsmath}

\usepackage{amsfonts}

\title{\bf Classical Characterization of quantum waves:\\
Comparison between the caustic and the zeros of the
Madelung-Bohm potential}

\author[1,*]{E. Esp\'indola-Ramos}
\author[1]{G. Silva-Ortigoza}
\author[1]{C. T. Sosa-S\'anchez}
\author[1]{I. Juli\'an-Mac\'ias}
\author[1]{ A. Gonz\'alez-Ju\'arez}
\author[1]{O. de J. Cabrera-Rosas}
\author[1]{P. Ortega-Vidals}
\author[2]{C. Rickenstorff-Parrao}
\author[3]{R. Silva-Ortigoza}

\affil[1]{\it Facultad de Ciencias F\'isico Matem\'aticas, Benem\'erita Universidad Aut\'onoma de Puebla. C.P. 72570, Puebla, Puebla, M\'exico}

\affil[2]{ \it Facultad de Ciencias de la Electr\'onica, Benem\'erita Universidad Aut\'onoma de Puebla. C.P. 72570, Puebla, Puebla, M\'exico}

\affil[3]{ \'Areas de Tecnolog\'ia de Computaci\'on Inteligente \& Mecatr\'onica, CIDETEC, Instituto Polit\'ecnico Nacional,  C.P. 07700, Mexico City, Mexico}

\affil[*]{Corresponding author: ernestoer@live.com.mx}

\date{}

\begin{document}


\maketitle


\begin{abstract}
\noindent From a geometric perspective, the caustic is the most classical description of a wavefunction since its evolution is governed by the Hamilton-Jacobi equation. On the other hand, according to the Madelung-de Broglie-Bohm equations, the most classical description of a solution to the Schr\"odinger equation is given by the zeros of the Madelung-Bohm potential. In this work, we compare these descriptions and, by analyzing how the rays are organized over the caustic, we find that the wavefunctions with fold caustic are the most classical beams because the zeros of the Madelung-Bohm potential coincide with the caustic. For another type of beams,  the Madelung-Bohm potential is in general distinct to zero over the caustic. We have verified these results for the one-dimensional Airy and Pearcey beams, which accordingly to the catastrophe theory, their caustics are stable. Finally, we remark that for certain cases, the zeros of the Madelung-Bohm potential are linked with the superoscillation phenomenon.
\end{abstract}


\section{Introduction}
\noindent The geometric description of a wavefunction is a practical tool to study  wave propagation in quantum mechanics and optics, and even in the general relativity background. The knowledge of the set of rays and geometrical wavefronts associated with the wavefunction enables us to study a series of properties and approximations such as self-reconstruction \cite{Bandres_2013,Voloch_Bloch_2013},  the classical description of the beam \cite{Berry_1979,Esp_ndola_Ramos_2018}, image formation, lens design and ronchigrams \cite{Berry_1987,2007,Bret_n_2017,Ju_rez_Reyes_2018}, and even an approximation for the field associated with the refraction phenomena \cite{Ortega_Vidals_2017}. Nowadays, the research on the engineering of structured beams has an astonishing impulse due to its applications \cite{Rubinsztein_Dunlop_2016,Zannotti_2020}, from nanoparticle  manipulation and nanotechnology applications \cite{Ashkin_1986,Grier_2003,Taylor_2015,Leach_2004,Capitanio_2005}  to the enhancement  of communication systems\cite{Gibson_2004,Bozinovic_2013,Krenn_2016}, high-resolution observation \cite{Vettenburg_2014,Yu_2015} and metrology\cite{Belmonte_2011,Rosales_Guzm_n_2013,Belmonte_2015}, to name a few. In all these cases the caustic (that is, the envelope of the rays) plays an important role since it does not only characterizes the region of maximum contribution to the intensity but also it determines the stability of the beam \cite{Nye_1979} and is the most classical description of the wavefunction since its evolution is governed by the Hamilton-Jacobi equation\cite{Esp_ndola_Ramos_2019}. On the other hand, the Madelung-de Broglie-Bohm equations of quantum mechanics also enables us to associate a classical description to a wavefunction through a modified Hamilton-Jacobi equation, which contains an additional potential (in comparison to the well known Hamilton-Jacobi equation) that we call as Madelung-Bohm potential\cite{Madelung_1927,Bohm_1952}. If the Madelung-Bohm potential is negligible in comparison to the ``classical energy'', then the macroscopic Hamilton-Jacobi equation is recovered. Nevertheless,  at the quantum level, the usual  Hamilton-Jacobi equation is also recovered over the space-time points where the Madelung-Bohm potential is equal to  zero. Thus the zeros of the Madelung-Bohm potential correspond to the most classical counterpart of a wavefunction. The aim of the present work is to compare the most classical description given by the geometrical approximation,  the caustic, with respect to the Madelung-de Broglie-Bohm equations, the zeros of the Madelung-Bohm potential, associated with a solution to the Schr\"odinger equation. To this end, first we present in sections II and III a brief review on the geometrical description of a wavefunction and the concept of Madelung-Bohm potential. Subsequently, in section IV, we introduce what we call {\it the interaction potential between plane waves}, which allow us to  deduce the conditions such that the caustic and the zeros of the Madelung-Bohm potential coincide.  Finally, in sections V and VI we focus on the Airy and Pearcey beams, which accordingly to the catastrophe theory, their caustics are stable (hence their importance). 

\section{The geometrical description of a wavefunction}

\noindent In a series of papers \cite{Cabrera_Rosas_2016,Sosa_S_nchez_2017,Juli_n_Mac_as_2018,Cabrera_Rosas_2019,Gonz_lez_Ju_rez_2019}, we describe the process to obtain the geometrical description of a wavefunction of the form
\begin{eqnarray}
\Phi(x,t)=\int_{\mathbb{R}} O(P) \exp\left(\frac{i}{\hbar} S(x,P,t)\right) dP.
\end{eqnarray}  
If $S(x,P,t)$ is a solution to both the Hamilton-Jacobi equation  and Laplace equation, then $\Phi(x,t)$ satisfies the Schr\"odinger equation for an arbitrary real function $O(P)$ \cite{Esp_ndola_Ramos_2018,Esp_ndola_Ramos_2019}. Under these conditions, the rays are determined from the stationary points of $S(x,P,t)$, that is to say
\begin{eqnarray}\label{Rays}
\frac{\partial S(x,P,t)}{\partial P}=0,
\end{eqnarray}
and the caustic is given by all the points that  satisfy the above equation and 
\begin{eqnarray}
\frac{\partial^{2} S(x,P,t)}{\partial P^{2}}=0. \label{Caustic}
\end{eqnarray}
The Hamilton's Principal Function associated to the wavefunction $\Phi(x,t)$, that we will denote by $\tilde{S}(x,t)$, is obtained as follows. From Eq.~(\ref{Rays}) one obtains locally
\begin{eqnarray}
P=P(x,t),
\end{eqnarray}
then we define
\begin{eqnarray}
\tilde{S}(x,t)\equiv S(x,P(x,t),t).\label{DCWD}
\end{eqnarray}
So then $\tilde{S}(x,t)$ defined in  Eq.~(\ref{DCWD}) satisfies the same Hamilton-Jacobi equation as $S(x,P,t)$ does. For this reason the geometrical wavefronts associated with $\Phi(x,t)$ are defined as the level curves of $\tilde{S}(x,t)$.  If the wavefunction propagates on free space, the set of singular points of the geometrical wavefronts is the caustic, which does not only characterizes the region of maximum contribution to the Probability Density Function but also determines the stability of the beam \cite{Nye_1979} and it represents the most classical description of the wavefunction $\Phi(x,t)$ since its evolution is governed by the Hamilton-Jacobi equation\cite{Esp_ndola_Ramos_2019}. It is worth noting that the caustic represents the region of transition on which the number of rays that intersect on each point of space-time, changes. This is directly deduced from  Eqs.~(\ref{Rays}) and (\ref{Caustic}) since the caustic is the envelope of the rays. \\

\noindent From now on, we restrict our attention to the cases
where $O(P)=constant$.

\section{A general overview of the Madelung-de Broglie-Bohm equations}

\noindent Any complex wavefunction can be represented on its polar form by
\begin{eqnarray}
\psi({\bf r},t)=R({\bf r},t)\exp\left(\frac{i}{\hbar} S_{MB}({\bf r},t) \right),
\end{eqnarray}
where $R({\bf r},t)$ and $S_{MB}({\bf r},t)$ are twice differentiable real functions. Consequently,  the Schr\"odinger equation is equivalent to a system of two equations obtained from its real and imaginary parts:
\begin{gather}
 \nabla \cdot \left(R^{2}({\bf r},t)\frac{\nabla S_{MB}({\bf r},t)}{m}\right)=\frac{\partial R^{2}({\bf r},t)}{\partial t},\label{Im(SE)}\\
 \frac{1}{2m}\left( \nabla S_{MB}({\bf r},t)\right)^{2} +V({\bf r }) +\frac{\partial S_{MB}({\bf r},t) }{\partial t} -\frac{\hbar^{2}}{2m}\frac{\nabla^{2}R({\bf r},t)}{R({\bf r},t)}=0.
 \label{Re(SE)}
 \end{gather}

\noindent Eq.~(\ref{Im(SE)}) is the continuity equation for the probability density $R^{2}({\bf r},t)$ with probability current density given by ${\bf J}=R^{2}({\bf r},t)\nabla S_{MB}({\bf r},t)/m$, whereas Eq.~(\ref{Re(SE)}) is sometimes called the Hamilton-Jacobi-Madelung equation.
This mathematical formulation was first introduced by Madelung in his hydrodynamical interpretation of quantum mechanics\cite{Madelung_1927}, and after rediscovered by David Bohm on his alternative interpretation of quantum mechanics in terms of hidden variables\cite{Bohm_1952}, where the idea of a pilot-wave driving the particle in a deterministic way,  proposed by de Broglie in 1927, is reconsidered. Throughout this paper, we will define the additional ``quantum term" in Eq.~(\ref{Re(SE)}) as the Madelung-Bohm potential $Q({\bf r},t)$:
\begin{eqnarray}
Q({\bf r},t)\equiv-\frac{\hbar^{2}}{2m}\frac{\nabla^{2}R({\bf r},t)}{R({\bf r},t)}.\label{QA}
\end{eqnarray}
From Eq.~(\ref{Re(SE)}), it is clear that the macroscopic Hamilton-Jacobi equation is recovered if the Madelung-Bohm potential is negligible with respect to the ``classical energy'' of the particle (a large value on the mass of the particle could allow this approximation as can be seen from Eq.~(\ref{QA})). Nevertheless, at the quantum level, over the space-time points, $({\bf r},t)$, where the  Madelung-Bohm potential is zero, the usual Hamilton-Jacobi equation is also recovered. In this sense, we say that the zeros of the Madelung-Bohm potential correspond to the most classical counterpart of a wavefunction, hence the importance of comparing them with the caustic region. On the other hand, from  Eq.~(\ref{QA}) we deduce that the zeros of the Madelung-Bohm potential are zero curvature points of the square root of the Probability Density Function (but not necessarily all of its zero curvature points due to the presence of the $R({\bf r},t)$ function in the denominator) for a fixed value of time $t$. This suggests that, in an analogous way to the caustic, the curves generated by the zeros of the Madelung-Bohm potential give a qualitative characterization of the Probability Density Function. 

\section{Quantum wave interactions in  vacuum}

\begin{figure*}[htbp]
\centering
\includegraphics[width=\linewidth]{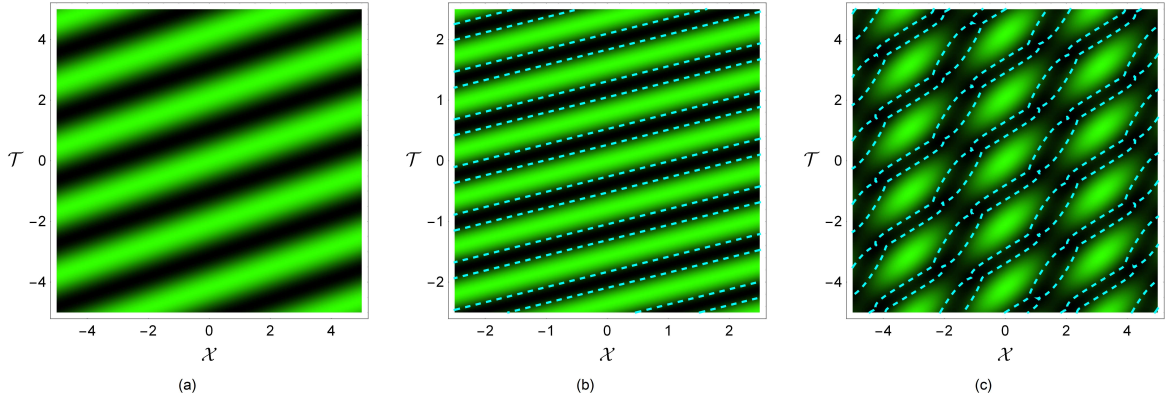}
\caption{\label{fig:epsart1} (a) Superposition of two plane waves with different momentum given by $\mathcal{P}=1$ and  $\mathcal{P}=2$. (b) Superposition of two waves with the same momentum given by  $\mathcal{P}=1$, with a third wave with momentum  $\mathcal{P}=3$. (c) Superposition of three different waves with momentum  $\mathcal{P}=1$,  $\mathcal{P}=-1$ and  $\mathcal{P}=2$. The dashed lines correspond to the zeros of the Madelung-Bohm potential. In all cases, for the Eq.~(\ref{PlaneWave})  we assume $f(P)=0$. For these graphs we have defined $\mathcal{X}=\sqrt{2}x/\hbar^{3/4}$, $\mathcal{T}=t/m\hbar^{1/2}$ and $\mathcal{P}=P/\sqrt{2}\hbar^{1/4}$. }
\end{figure*}

\noindent The wavefunction associated with the one-dimensional free particle with classical momentum $P$ is given by 
\begin{eqnarray}
\psi_{(P)} (x,t)=\exp \left( \frac{i}{\hbar} S_f(x,P,t)\right), \label{PlaneWave}
\end{eqnarray}
where
\begin{eqnarray} 
S_f(x,P,t)= xP-\frac{P^{2}}{2m}t+f(P), 
\end{eqnarray}
satisfies the Hamilton-Jacobi equation for a free particle and the Laplace equation, and $f(P)$ is an arbitrary twice differentiable real function. From the  Eqs.~(\ref{QA}) and (\ref{PlaneWave}), it is clear that the Madelung-Bohm potential associated with a one-dimensional plane wave is zero. On the other hand, the wavefunction associated with the superposition of two plane waves with momentum $P$ and $P+\Delta P$ is given by
\begin{eqnarray}
\psi_{(P)} (x,t) + \psi_{(P+\Delta P)} (x,t)= \psi_{(P)} (x,t) \left[ 1+\exp\left(\frac{i}{\hbar}S_{\Delta P}(x,P,t)\right)\right],\label{SumTW}
\end{eqnarray}
where we have defined $S_{\Delta P}$ as
\begin{eqnarray}
S_{\Delta P}(x,P,t)\equiv x\Delta P-\left(\frac{(\Delta P)^{2}}{2m}+\frac{P\Delta P}{m}\right)t+f(P+\Delta P)-f(P). \label{DP}
\end{eqnarray}
According to  Eq.~(\ref{QA}) the Madelung-Bohm potential associated with Eq.~(\ref{SumTW}) is \begin{eqnarray}
Q_{I}(x,t)=\frac{(\Delta P)^{2}}{8m},\label{QIP1}
\end{eqnarray}
which is constant. We define the above result as the interaction potential $Q_{I}(x,t) $ of two quantum plane waves. If both waves have the same momentum,  $\Delta P =0$, then $Q_{I}(x,t) =0$, which is equivalent to having a single quantum wave, so there are no self-interactions. In figure \ref{fig:epsart1}(a) we illustrate the Probability Density Function for the superposition of two plane waves.\\
\\
For the superposition of three plane waves, the Madelung-Bohm potential has a long expression, as shown in the appendix. In this case, the zeros of the potential are one-dimensional curves surrounding the maxima of the intensity pattern as shown in figures \ref{fig:epsart1}(b) and \ref{fig:epsart1}(c). Thus the interaction potential $Q_{I}(x,t) $  of the three waves depends on space-time coordinates.\\ 

\begin{figure*}[htbp]
\centering
\includegraphics[width=\linewidth]{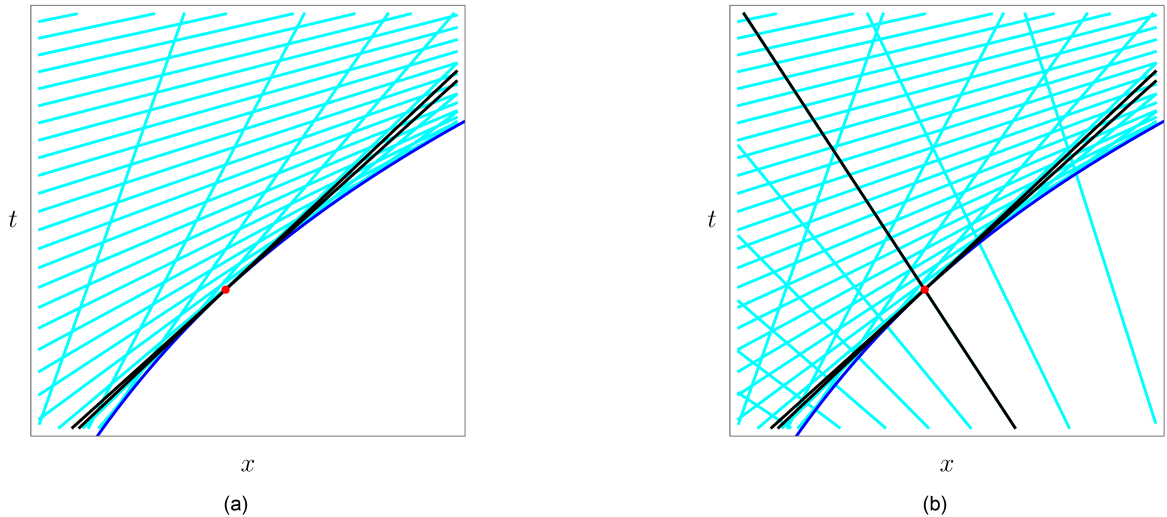}
\caption{\label{fig:epsart2} (a) A beam for which two rays pass at each point on its caustic in a tangent way. (b) A beam for which exist a third ray (or even more) on each point of the caustic crossing it. In both figures, the caustic is the blue curve.
 }
\end{figure*}

\noindent Let us now consider the wavefunction $\Phi(x,t)$ built as the infinite superposition of plane waves as follows
\begin{eqnarray}
\label{SumInt}
\Phi (x,t)=\int_{-\infty}^{\infty} \exp\left(\frac{i}{\hbar}S_{f}(x,P',t)\right)dP' =\lim_{\Delta P \rightarrow 0} \Delta P \sum_{k=-\infty}^{\infty} \psi_{(P+k\Delta P)}(x,t),
\end{eqnarray}
where $P$ has an arbitrary real fixed value. By applying  Eq.~(\ref{SumTW}), the sum in the above equation can be rewritten in the following manner 
\begin{multline}\label{limit}
\sum_{k=-\infty}^{\infty} \psi_{(P+ k\Delta P)} (x,t) =  \lim_{ k \rightarrow -\infty}  \psi_{(P+ k\Delta P)}(x,t)\\+\sum_{k=-\infty}^{\infty} \psi_{(P+ k\Delta P)}(x,t)\exp\left(\frac{i}{\hbar}S_{\Delta P}(x,P+k\Delta P,t)\right).
\end{multline}
That is to say, every element of the sum corresponds to the superposition of two ``close" plane waves with momentum $ P+k\Delta P$ and $P+(k+1)\Delta P$. On the other hand, by assuming  that $\Delta P <<1$ (in order to get a good approximation of  Eq.~(\ref{SumInt}))
, we can approximate  Eq.~(\ref{DP}) to  
\begin{eqnarray}\label{spa}
S_{\Delta P}(x,P+k\Delta P,t) =  \Delta P \left(  \frac{\partial S_{f}(x,\alpha,t) }{\partial \alpha }\right)_{\alpha=P+k\Delta P}.
\end{eqnarray}
Then, the wavefunction $\Phi$ can be approximated to 
\begin{eqnarray}
\label{RayApprox}
\Phi (x, t  )  \approx \Delta P \sum_{k=-\infty}^{\infty} \psi_{(P+k\Delta P)}(x,t)\exp\left[\frac{i\Delta P}{\hbar} \left(  \frac{\partial S_{f}(x,\alpha,t) }{\partial \alpha }\right)_{\alpha=P+\Delta P}\right],
\end{eqnarray}
where the first term on the right side of Eq.~(\ref{limit}) is negligible in comparison to the sum, due to its value lies on the unitary circle of the complex plane (see Eq.~(\ref{PlaneWave})), and the equality is satisfied at the limit $\Delta P\rightarrow 0$. It is well known that for the wavefunction  $\Phi(x,t)$ the condition $\partial S_f(x,P,t) / \partial P =0$ determines the rays or the  semiclassical trajectories associated to the beam. The last two equations suggest that the notion of rays emerge when the plane wavefronts are very close to each other with respect to their momentum. Then, at the limit $\Delta P \rightarrow 0$,  two closely rays pass to be infinitesimally adjacent, and their intersection point (if exist) becomes part of the caustic of the beam. In what follows we restrict ourselves to the study of beams with caustic. At this point it is important to distinguish the two fundamental forms on which the rays are organized over the caustic during  propagation. In the first case, at each point of the caustic pass only two tangent rays as shown in figure \ref{fig:epsart2}(a); it is clear that this condition is satisfied by a fold caustic (see, for example \cite{Berry_1980}). For the second case, there exist a third ray (or even more) that crosses the caustic at each point. That is to say, at each point of the caustic there are two tangent rays and one or more that cross the caustic at the same point, as it is shown in figure \ref{fig:epsart2}(b).
\\
\noindent It is important to remark that in the ondulatory approach, caustics are regions where the wavefunction changes from an oscillatory behavior, due to a strong interference effect, to an exponential decay in the ``shadow" region of the Probability Density Function, due to an evanescent field. In the geometric approach, caustics are curves where light rays are focused,  and then, characterize the region of maximum contribution to the Probability Density Function\cite{nye2000natural,Kravtsov_1999,Vaveliuk_2015}. Thus, a ray analysis near the caustic could enable us to elucidate some properties about the Probability Density Function, as we will see next.\\
\noindent  In the first case, over each point of the caustic pass two parallel rays associated with two parallel plane waves. Then,  according to Eq.~(\ref{QIP1}),   the  Madelung-Bohm potential is zero over the caustic, and the  Hamilton-Jacobi equation is recovered at those points. Thus, the wavefunctions with this type of caustic are the most classical beams,  because the evolution of the caustic, and then the evolution of the zeros of the Madelung-Bohm potential,  are governed by the Hamilton-Jacobi equation \cite{Esp_ndola_Ramos_2019}. Furthermore, the evolution in time of the maxima and minima points of the Probability Density Function travels with the caustic, forming a {\it uniform interference pattern} on the space-time framework (see, for example, figure \ref{fig:epsart3}). In fact, the Probability Density Function profile at the vicinity of the global maximum is always shaped like an Airy function, since it manifests an optical fold catastrophe \cite{Greenfield_2011}.\\
For the second case, due to the existence of a third ray (or even more),  the  Madelung-Bohm potential, in general, is not zero over the caustic, and unlike the first case,  the Probability Density Function has a {\it non-uniform interference pattern} in the sense that their maxima and minima points do not travel with the caustic during its time evolution  (see for example, figure \ref{fig:epsart4}). Therefore, strictly speaking, {\it there is no classical description for the non-uniform interference pattern}. 
\\
\noindent In what follows, we will analyze an example for each case studied in this section: the one-dimensional Airy and Pearcey beams, which accordingly to the catastrophe theory, their caustics are stable.

\section{The Airy beam}

\begin{figure*}[htbp]
\centering
\includegraphics[width=\linewidth]{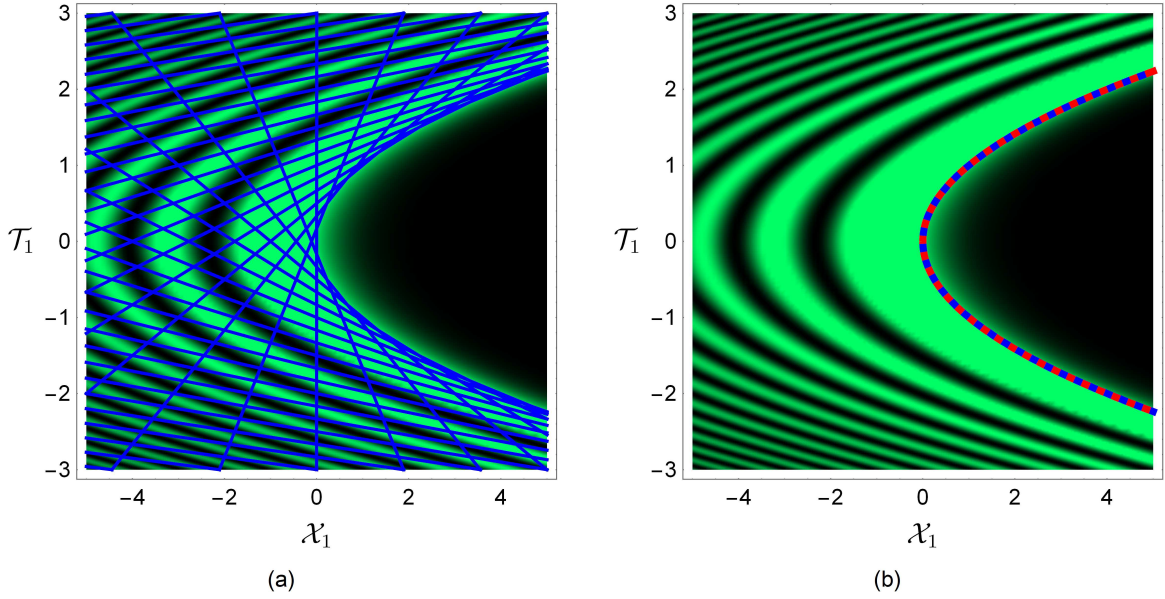}
\caption{\label{fig:epsart3} (a) The Probability Density Function of the Airy beam and its associated rays. (b) The zeros of the Madelung-Bohm potential (dashed red lines) and the caustic (blue curve). For these graphs we have defined $\mathcal{X}_{1}=B_{1}x/\hbar^{2/3}$, $\mathcal{T}_{1}=B_{1}^2 t/2m\hbar^{1/3}$ and $\mathcal{P}_{1}=P/B_{1}\hbar^{1/3}$ .}
\end{figure*}

\noindent The wavefunction associated with the one-dimensional Airy beam is given by \cite{Berry_1979}
\begin{eqnarray}
\Phi_A(x,t)=Ai\left[ \frac{B_{1}}{\hbar^{2/3}}\left( x-\frac{B_{1}^{3}t^{2}}{4m^{2}}\right)\right]\exp\left[i\left(\frac{B_{1}^{3}tx}{2m\hbar} -\frac{B_{1}^{6}t^{3}}{12m^{3}\hbar} \right)  \right],
\label{WAB}
\end{eqnarray}
where $B_{1}$ is an arbitrary positive constant with units $(kg^{2}m/s^{2})^{1/3}$, and $Ai$ denotes the Airy function. Thus, according to Eq.~(\ref{QA}), the Madelung-Bohm potential is
\begin{eqnarray}
Q(x,t)=\frac{B_{1}^{6} t^{2}-4B_{1}^{3}m^{2}x}{8m^{3}}. \label{QPABPRA}
\end{eqnarray}
On the other hand, the Airy beam has the following decomposition in terms of plane waves 
\begin{eqnarray}
\Phi_A(x,t)= \frac{1}{2\pi \hbar^{1/3} B_{1}} \int_{-\infty}^{\infty} \exp\left[\frac{i}{\hbar}\left( xP-\frac{P^{2}}{2m}t+\frac{P^{3}}{3B_{1}^{3}} \right) \right]dP.
\label{AIF}
\end{eqnarray}
Therefore for this beam 
\begin{eqnarray}
S_{f}(x,P,t)=xP-\frac{P^{2}}{2m}t+\frac{P^{3}}{3B_{1}^{3}},
\end{eqnarray}
from the stationary points of $S_f(x,P,t)$ we obtain 
\begin{eqnarray}
P(x,t)=\frac{B_{1}^{3}t}{2m}\pm\frac{\sqrt{B_{1}^{6} t^{2}-4B_{1}^{3}m^{2}x}}{2m},\label{SPAB}
\end{eqnarray}
and the semiclassical description of the beam is given by 
\begin{eqnarray}
\tilde{S}_{f}  (x,t) = S_f  (x,P(x,t),t) =\left(\frac{B_{1}^{3}tx}{2m}-\frac{B_{1}^{6}t^{3}}{12m^{3}}\right) \mp \frac{B_{1}^{3/2}}{12m^{3}}\left( B_{1}^{3} t^{2}-4m^{2}x\right)^{3/2}.\label{SDABPRA}
\end{eqnarray}
The caustic is given by 
\begin{eqnarray}\label{AiC}
B_{1}^{6} t^{2}-4B_{1}^{3}m^{2}x=0,
\end{eqnarray}
which is directly deduced from  Eq.~(\ref{SPAB}),   since the caustic is the region of transition from  zero  to two rays. From  Eqs.~(\ref{QPABPRA}) and (\ref{AiC}) we can check that the Madelung-Bohm potential is zero over the caustic. In figure \ref{fig:epsart3} we illustrate the Probability Density Function, the rays, the caustic and the zeros of the Madelung-Bohm potential associated with this beam.  \\
\\
It is important to remark that, for this example, the physical phase of the beam can be determined from a geometric perspective. From Eq.~(\ref{SDABPRA}), notice that the first term gives the physical phase of the beam times $\hbar$, whereas the second term is zero over the caustic. This fact has its origin on the integral  Eq.~(\ref{AIF}) due to the change of variable
\begin{eqnarray}
U=P-\frac{B_{1}^{3}t}{2m},\label{ChV}
\end{eqnarray}
where the last term can be deduced from the first term of Eq.~(\ref{SPAB}) since $P(x,t)$ is single-valued over the caustic, enable to transform the integral representation of the beam to its polar form as shown in  Eq.~(\ref{WAB}). Then, the physical phase can be calculated as follows
\begin{eqnarray}
\frac{S_{f}(x,P-U,t)}{\hbar}=\frac{B_{1}^{3}tx}{2m\hbar} -\frac{B_{1}^{6}t^{3}}{12m^{3}\hbar}. \label{PFFGF}
\end{eqnarray}
It is well known that near a fold caustic the wavefunction is approximated to an Airy function, which is real valued. Thus there must exist beams for which the physical phase is given by  Eq.~(\ref{PFFGF}). All these results can also be verified for the nondiffracting Bessel beam, which has a fold caustic. In fact, a direct computation shows that the caustic and the zeros of the Madelung-Bohm potential coincide, separating the region where the superoscillation phenomenon occurs from where it is not \cite{Cabrera_Rosas_2016,Berry_2013,berry2020superoscillations}.\\
\\
In recent works,  Esp\'indola-Ramos et al.\cite{Esp_ndola_Ramos_2018}, Gonz\'alez-Ju\'arez and Silva-Ortigoza \cite{Gonz_lez_Ju_rez_2019}, showed that the Airy and Bessel beams can be completely described from a geometrical perspective (or semiclassical perspective). The Hamilton-Jacobi theory suggests a generalization for the rays Eq.~(\ref{Rays}) given by
\begin{eqnarray}
\frac{\partial S(x,P,t)}{\partial P}=\mathcal{L},
\end{eqnarray}
where $\mathcal{L}$ is a real parameter that enables us to generate a family of caustics which characterize all the maxima and minima of the Probability Density Function. This fact could be intimately linked with the result of the Eq.~(\ref{PFFGF}): the fact that the geometrical wavefronts and the physical wavefronts coincide at the caustic region would allow constructing a complete geometrical description of these kind of beams.

\section{The Pearcey beam}

\begin{figure*}[htbp]
\centering
\includegraphics[width=\linewidth]{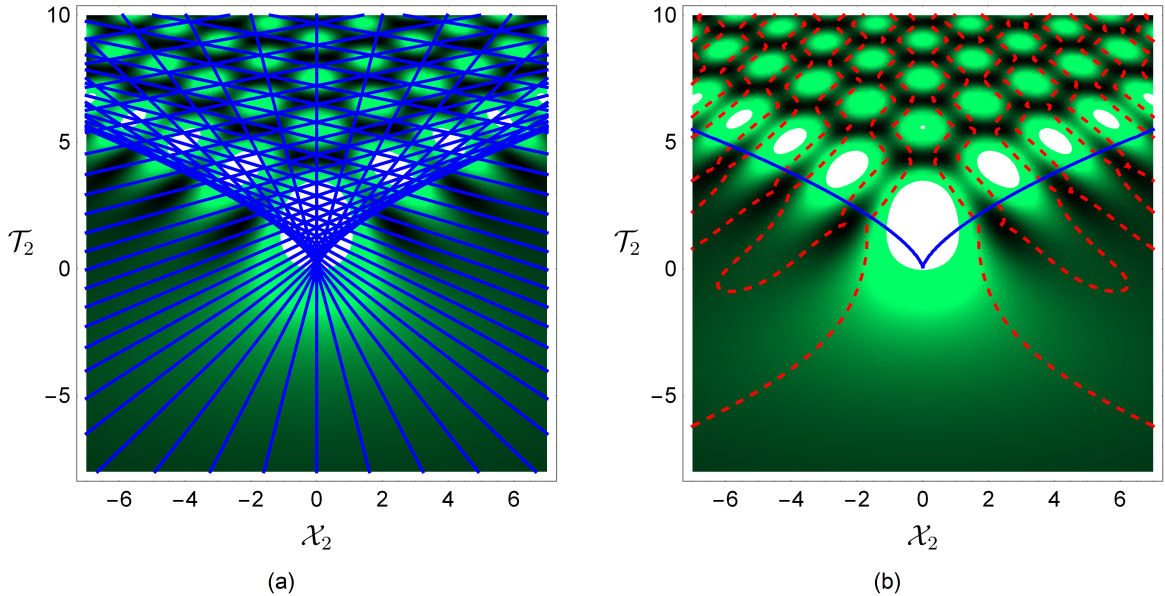}
\caption{\label{fig:epsart4}(a) The Probability Density Function of the Pearcey beam and its associated rays. (b) The zeros of the Madelung-Bohm potential (dashed red lines) and the caustic (blue curve). For these graphs we have defined $\mathcal{X}_{2}=\sqrt{2}B_{2}x/\hbar^{3/4}$, $\mathcal{T}_{2}=B_{2}^2 t /m\hbar^{1/2}$ and  $\mathcal{P}_{2}=P / \sqrt{2} B_{2}\hbar^{1/4}$.}
\end{figure*}

\noindent The one-dimensional Pearcey beam is defined as the infinite superposition of plane waves in the following manner  
\begin{eqnarray}
\Phi_{Pe}(x,t)=\int_{-\infty}^{\infty}\exp\left[\frac{i}{\hbar}\left(   xP-\frac{P^{2}}{2m}t+\frac{P^{4}}{4 B_{2}^4} \right)   \right]dP,
\end{eqnarray}
where $B_{2}$ is an arbitrary positive constant with units $(kg^3 m^2 /s^3)^{1/4}$. We identify $S_{f}(x,P,t)$ as
\begin{eqnarray}
S_{f}(x,P,t)=xP-\frac{P^{2}}{2m}t+\frac{P^{4}}{4B_{2}^4},
\end{eqnarray}
and the stationary points of $S_{f}(x,P,t)$ must satisfy  
\begin{eqnarray}
\mathcal{P}_{2}^3-\frac{\mathcal{P}_{2}^2 \mathcal{T}_{2}}{2}+\frac{\mathcal{X}_{2}}{4}=0,\label{SPPB}
\end{eqnarray}
 where we have defined
\begin{eqnarray}
\mathcal{X}_{2}=\frac{\sqrt{2}B_{2}x}{\hbar^{3/4}},\quad \mathcal{T}_{2}=\frac{B_{2}^{2}t}{m\hbar^{1/2}},\quad \mathcal{P}_{2}=\frac{P}{\sqrt{2}B_{2}\hbar^{1/4}}.
\end{eqnarray}
According with the formulas of Cardano, the sign of $\Delta(\mathcal{X}_{2},\mathcal{T}_{2})$,  given by
\begin{eqnarray}
\Delta(\mathcal{X}_{2},\mathcal{T}_{2})=\frac{1}{1728}\left( 27\mathcal{X}_{2}^2-8\mathcal{T}_{2}^3\right),
\end{eqnarray}
determines the number and the multiplicity of the real roots for $\mathcal{P}_{2}(\mathcal{X}_{2},\mathcal{T}_{2})$ which satisfy Eq.~(\ref{SPPB}). Thus, it is clear that the caustic is determined by the condition $\Delta(\mathcal{X}_{2},\mathcal{T}_{2})=0$, or
\begin{eqnarray}
27\mathcal{X}_{2}^{2}-8\mathcal{T}_{2}^{3}=0.
\end{eqnarray}
If $\Delta(\mathcal{X}_{2},\mathcal{T}_{2})>0$, there is a single real root  given by
\begin{eqnarray}
\mathcal{P}_{2}(\mathcal{X}_{2},\mathcal{T}_{2})=\sum_{n=0}^{1}\left( -\frac{\mathcal{X}_{2}}{8}-(-1)^{n}\sqrt{\Delta(\mathcal{X}_{2},\mathcal{T}_{2})}\right)^{1/3}.
\end{eqnarray}
Then, at each point outside the caustic, pass a single ray as is shown in figure \ref{fig:epsart4}(a) (at the lower side of the envelope of the rays).\\
If $\Delta(\mathcal{X}_{2},\mathcal{T}_{2}) < 0 $, there are three real roots given by
\begin{eqnarray}
\mathcal{P}_{2}(\mathcal{X}_{2},\mathcal{T}_{2})=2\sqrt{\frac{\mathcal{T}_{2}}{6}}\cos\left(  \frac{\theta +2k\pi}{3}  \right),
\end{eqnarray}
where $k=0,1,2$ and  $0<\theta<\pi$ must satisfy
\begin{eqnarray}
\cos\theta =-\frac{\mathcal{X}_{2}}{8\sqrt{(\mathcal{T}_{2}/6)^3}}.
\end{eqnarray}
That is to say, there are three rays that intersect over each point inside the caustic (see figure \ref{fig:epsart4}(a)).\\
If $\Delta(\mathcal{X}_{2},\mathcal{T}_{2})=0$, at the singularity of the caustic, $(\mathcal{X}_{2},\mathcal{T}_{2})=(0,0)$, there is a  real root of multiplicity three given by
\begin{eqnarray}
\mathcal{P}_{2}(0,0)=0.
\end{eqnarray}
On the other hand, over the caustic there are three real roots. One of them has multiplicity two and it is given by 
\begin{eqnarray}
\mathcal{P}_{2}(\mathcal{X}_{2},\mathcal{T}_{2})=\frac{3\mathcal{X}_{2}}{4\mathcal{T}_{2}}.
\end{eqnarray}
and the other root is given by
\begin{eqnarray}
\mathcal{P}_{2}(\mathcal{X}_{2},\mathcal{T}_{2})=-\frac{4\mathcal{T}_{2}^2}{9\mathcal{X}_{2}}.
\end{eqnarray}
Thus, at each point over the caustic (except at its singularity), pass two parallel rays that intersect a third one that crosses the caustic.\\
\\
Notice that the existence of a third ray that crosses the caustic in non-tangent way gives rise to a {\it non-uniform interference pattern} for the Probability Density function, as we can see in figure \ref{fig:epsart4}(a).  In figure \ref{fig:epsart4}(b) we illustrate the caustic and the zeros of the Madelung-Bohm potential, which have been numerically calculated by using the Connor representation of Pearcey integral (see for example \cite{L_pez_2016}). In contrast to the Airy beam, the geometry of the caustic and the zeros of the Madelung-Bohm potential are completely different. While the caustic characterize the distribution of the lobes with the maximum contribution to the Probability Density Function, the zeros of the Madelung-Bohm potential surround the maxima of the interference pattern, giving us a better idea of the structure of the Probability density function, {\it possibly a factor to take into account in the design of structured beams}.  Consequently, strictly speaking, the Pearcey interference pattern near to the caustic has no counterpart in classical mechanics.

\section{Conclusions}

\noindent While the caustic corresponds to the most classical counterpart of a wavefunction from a geometric perspective since its evolution is governed by the Hamilton-Jacobi equation \cite{Esp_ndola_Ramos_2019}, the Madelung-de Broglie-Bohm equations also enables us to find the most classical regions of the wavefunction defined by the zeros of the Madelung-Bohm potential, in the sense that the Hamilton-Jacobi equation is recovered at those points. By introducing the concept of the interaction potential between plane waves and analyzing how the rays are organized over the caustic, we conclude that the wavefunctions with fold caustic are the most classical beams because the curve generated by the zeros of the Madelung-Bohm potential coincides with the caustic, and therefore its evolution also is governed by the Hamilton-Jacobi equation.  We have verified this fact for the one-dimensional Airy beam. Additionally, we find that the physical phase for the Airy beam can be obtained from a geometric perspective since the physical phase of the beam and the geometrical wavefronts coincide over the caustic. We conjecture that these results are linked with the observations made by  Esp\'indola-Ramos et al.\cite{Esp_ndola_Ramos_2018}, Gonz\'alez-Ju\'arez and Silva-Ortigoza \cite{Gonz_lez_Ju_rez_2019}, for the Airy and Bessel beams; they found a family of caustics which characterize in qualitative and quantitative form all the maxima of the Probability Density Function. On the other hand, we find that if there are rays that cross the caustic in a non-tangent form, then a {\it non-uniform interference pattern} is generated near to the caustic, so that the caustic and the zeros curves of the Madelung-Bohm potential are completely different. Hence, strictly speaking, {\it there is no counterpart in classical mechanics for the non-uniform interference pattern}. We have verified these facts for the one-dimensional Pearcey beam.\\
\\
\noindent It is important to remark that the Withney's theorem\cite{Arnold_2012} implies that the fold and the cusp are the only locally stable singularities in the space-time $(x,t)$. Thus any unstable beam under a small perturbation leads to one of the cases studied here. It could be interesting to extend our study for the stable caustics in higher dimensions. In those cases, the zeros of the Madelung-Bohm potential will be two-dimensional surfaces or hypersurfaces. \\
\\
It is well known that when the geometrical approach is valid for the description of a beam, the caustic characterizes the maxima of the Probability Density Function. On the other hand, the pattern of phase singularities also gives an alternative characterization of the Probability Density Function, with the advantage of being complementary to the caustic\cite{1974}, in the sense that it indicates perfect destructive interference -the ultimate nonclassical feature. We believe that the geometry of the zeros of the Madelung-Bhom potential also gives a  characterization of the structure of the Probability Density Function since it corresponds to the zero curvature points of the square root of the Probability Density Function. For the ``most classical beams'' this geometry coincides with the caustic, and for another type of beams, the zeros of the Madelung-Bohm potential surround the maxima of the interference pattern. We think that these properties could be taken into account for the design of structured beams.  \\
\\
After finishing this work, M. V. Berry remarked us that the zeros of the Madelung-Bohm potential separate regions where the (band-limited by the wavenumber ) wave is superoscillatory (i.e. varying faster than any of the plane waves in its superposition \cite{Berry_2006,Berry_2008,Berry_2013,Berry_2019}), from where is not, thus linking the large literature about superoscillations with the large literature on Madelung-Bohm. In a recent paper, he studies this relationship\cite{berry2020superoscillations}.

\section*{Appendix. The Madelung-Bohm potential for three interacting plane waves}

\noindent Let's consider the superposition of three waves with momentum $P$, $ P+\Delta P$ and $P+\Delta P'$. We will denote the corresponding wavefunction as $\psi_{3}(x,t)$, and is given by 
\begin{eqnarray}
\psi_3(x,t) =\psi_{(P)}(x,t)+\psi_{(P+\Delta P)}(x,t)+\psi_{(P+\Delta P')}(x,t).
\end{eqnarray}
By using Eq.~(\ref{SumTW}), we can rewrite the above equation as follows 
\begin{eqnarray}
\psi_3 (x,t) = \psi_{(P)}(x,t) \left[ 1+\exp\left(\frac{i}{\hbar}S_{\Delta P}(x,P,t)\right)
+\exp\left(\frac{i}{\hbar}S_{\Delta P'}(x,P,t)\right)\right].
\end{eqnarray}
We will denote the squared modulus of the wavefunction $\psi_{3}(x,t)$ as $R^2 (x,t)$, which, by applying trigonometric identities, turns out to be the following
\begin{eqnarray}
R^{2}(x,t)=1+8\cos\left(\frac{S_{\Delta P}}{2\hbar} \right)\cos\left(\frac{S_{\Delta P'}}{2\hbar} \right)\cos\left(\frac{S_{\Delta P}-S_{\Delta P'}}{2\hbar} \right).
\label{AppendixR2}
\end{eqnarray}
Therefore, according to Eq.~(\ref{QA}), the interaction potential, $Q_{I}(x,t)$, for three quantum plane waves is given by 
\begin{eqnarray}
Q_{I}(x,t)=\frac{A(x,t)R^{2}(x,t)-B(x,t)^{2}}{2m R^{4} (x,t)},
\end{eqnarray}
where $A(x,t)$ and $B(x,t)$ are given by

\begin{multline}
A(x,t)=(\Delta P)^{2} \cos  \left(\frac{S_{\Delta P}(x,P,t)}{\hbar}  \right)\\
+(\Delta P')^{2}\cos\left(\frac{S_{\Delta P'}(x,P,t)}{\hbar}\right)+ (\Delta P-\Delta P')^{2}\cos\left(\frac{\Delta'(x,P,t)}{\hbar}\right),
\end{multline}
and

\begin{multline}
B(x,t)=(\Delta P)  \sin  \left(\frac{S_{\Delta P}(x,P,t)}{\hbar}\right)\\
+(\Delta P')\sin\left(\frac{S_{\Delta P'}(x,P,t)}{\hbar}\right)
+(\Delta P-\Delta P')\sin\left(\frac{\Delta'(x,P,t)}{\hbar}\right),
\end{multline}

\noindent where we have defined $\Delta'(x,P,t)$ as
\begin{eqnarray}
\Delta'(x,P,t)=S_{\Delta P}(x,P,t)-S_{\Delta P'}(x,P,t).
\end{eqnarray}

\section*{Funding}
Sistema Nacional de Investigadores (SNI); VIEP- Benemérita Universidad Autónoma de Puebla (BUAP).

\section*{Acknowledgments}
The authors acknowledge the comments from M. V. Berry. Ernesto Esp\'indola Ramos, Israel Juli\'an Mac\'ias, Citlalli Teresa Sosa S\'anchez, and Adriana Gonz\'alez Ju\'arez were supported by a CONACyT scholarship. Gilberto Silva Ortigoza and Ram\'on Silva Ortigoza acknowledge financial support from Sistema Nacional de Investigadores (SNI). This work has received partial support fromVIEP-BUAP.

 \bibliographystyle{unsrt}
\bibliography{sample}


\end{document}